\documentclass{article}

\usepackage[final,nonatbib]{nips_2017}

\usepackage[utf8]{inputenc} 
\usepackage[T1]{fontenc}    
\usepackage{hyperref}       
\usepackage{url}            
\usepackage{booktabs}       
\usepackage{amsfonts}       
\usepackage{nicefrac}       
\usepackage{microtype}      
\usepackage{graphicx}
\usepackage{multirow}

\title{Raw Waveform-based Audio Classification Using Sample-level CNN Architectures}

\author{
  Jongpil Lee \\
  Graduate School of Culture Technology \\
  KAIST\\
  \texttt{richter@kaist.ac.kr} \\
   \And
   Taejun Kim \\
   School of Electrical and Computer Engineering \\
   University of Seoul \\
   \texttt{ktj7147@uos.ac.kr} \\
   \AND
   Jiyoung Park \\
   Graduate School of Culture Technology \\
   KAIST \\
   \texttt{jypark527@kaist.ac.kr} \\
   \And
   Juhan Nam \\
   Graduate School of Culture Technology \\
   KAIST \\
   \texttt{juhannam@kaist.ac.kr} \\
}

\begin{document}

\maketitle

\begin{abstract}

 Music, speech, and acoustic scene sound are often handled separately in the audio domain because of their different signal characteristics. However,  as the image domain grows rapidly by versatile image classification models, it is necessary to study extensible classification models in the audio domain as well. In this study,  we approach this problem using two types of sample-level deep convolutional neural networks that take raw waveforms as input and uses filters with small granularity. One is a basic model that consists of convolution and pooling layers. The other is an improved model that additionally has residual connections, squeeze-and-excitation modules and multi-level concatenation.  
We show that the sample-level models reach state-of-the-art performance levels for the three different categories of sound. 
Also, we visualize the filters along layers and compare the characteristics of learned filters.
\end{abstract}

\section{Introduction}
Broadly speaking, audio classification tasks are divided into three sub-domains including music classification, speech recognition (particularly for the acoustic model), and  acoustic scene classification. However, input audio features and models for each sub-domain task are usually different due to the different signal characteristics. 


Recent advances in deep learning have encouraged a single audio classification model to be applied to many cross-domain tasks.  For example, Adavanne et. al. used a Convolutional Recurrent Neural Network (CRNN) model for sound event detection \cite{adavanne2017sound}, bird audio classification \cite{adavanne2017stacked} and music emotion recognition \cite{malik2017stacked}. 
However, depending on the task, the majority of the audio classification models use different sub-optimal settings of time-frequency representation as input in terms of filter-bank type and size, time-frequency resolution and magnitude compression. This  in turn influences model architecture, for example, the choices of convolutional layer (1D or 2D) and filter shape \cite{pons2016experimenting}.

This issue can be solved by a waveform-based model that directly takes raw input signals.
Recently, Dieleman and Schrauwen used raw waveforms as input of CNN models for music auto-tagging task \cite{dieleman2014end}. Sainath et. al. used Convolutional Long short-term memory Deep Neural Network (CLDNN) for speech recognition \cite{sainath2015learning}. Dai et. al. used Deep Convolutional Neural Networks (DCNN) with residual connections for environmental sound recognition \cite{dai2017very}. All of them used frame-level filters (typically several hundred samples long) in the first convolutional layer which were carefully configured to handle the target task. In this frame-level raw waveform input, however, the filters in the bottom layer should learn all possible phase variation of (pseudo-)periodic waveforms which are likely to be prevalent in audio signals. This has impeded the use of raw waveform as input over spectrogram-based representations where the phase variation within a frame (i.e. time shift of periodic waveforms) is removed by taking the magnitude only.  



The phase invariance is analogous to translation invariance in the image domain. Considering that the filter size is typically small in the image domain, even 3$\times$3 in the VGG model \cite{simonyan2014very}, we investigated the possibility of stacking very small-size filters from the bottom layer of DCNN for raw audio waveforms with max-pooling layers. 
The results on music auto-tagging \cite{lee2017sample} and sound event detection \cite{lee2017dcase} showed that the VGG-style 1-D CNN models are highly effective. We term this model as \textbf{SampleCNN}. 

We enhanced the SampleCNN model by adding residual connections, squeeze-and-excitation modules and multi-level feature concatenation for music auto-tagging \cite{kim2017rese}. The residual connection makes gradient propagation more fluent, allowing training deeper networks \cite{he2016deep}. The Squeeze-and-Excitation (SE) module recalibrates filter-wise feature responses \cite{hu2017squeeze}. The multi-level feature concatenation takes different abstraction levels of classification labels into account \cite{lee2017multi}. We term this model as \textbf{ReSE-2-Multi}. 




In this study, we show that the sample-level CNN models are effective for three datasets from different audio domains.
Furthermore, we visualize hierarchically learned filters for each dataset in the waveform-based model to explain how they process sound differently.


\section{Models}

\begin{figure}[t]
\vspace{1.6 mm}
  \centering
  \centerline{\includegraphics[width=\textwidth]{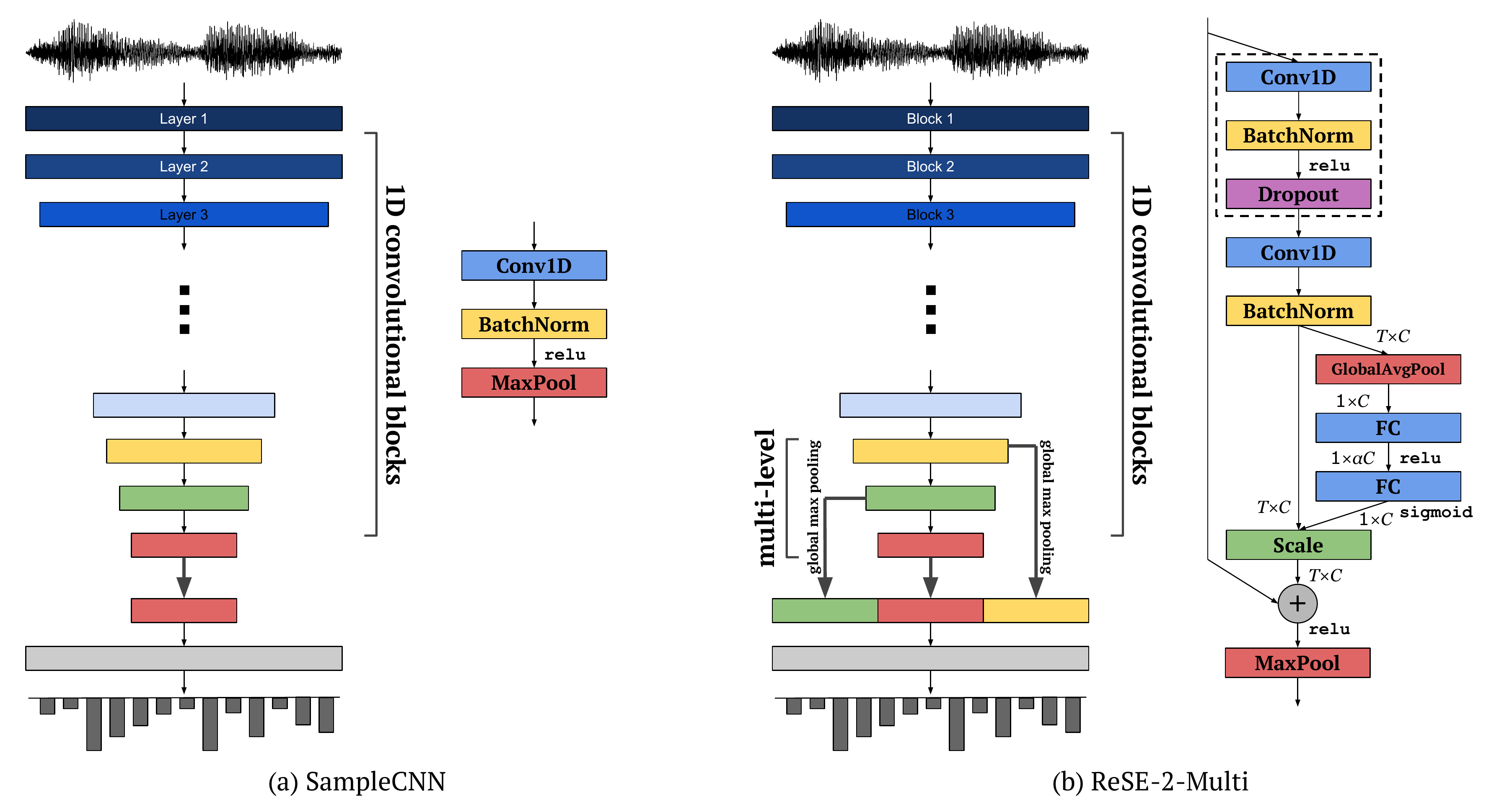}}
  \caption{SampleCNN and ReSE-2-Multi models.}
  \label{figure0}
\end{figure}


Figure \ref{figure0} shows the structures of the two sample-level models. 2 or 3 sample-size 1D filters and poolings are used in all convolutional layers. In SampleCNN, convolutional layer, batch normalization layer and max-pooling layer are stacked as shown in Figure \ref{figure0} (a). The detailed description can be found in \cite{lee2017sample}.

In ReSE-2-Multi, we add a residual connection and an SE module onto the SampleCNN building block as shown in Figure \ref{figure0} (b). The SE path recalibrates feature maps through two operations. One is squeeze operation that aggregates a global temporal information into filter-wise statistics using global average pooling. The operation reduces the temporal dimensionality ($T$) to one. The other is excitation operation that adaptively recalibrates each filter map using the filter-wise statistics from the squeeze operation and a simple gating mechanism. The gating consists of two fully-connected (FC) layers that compute nonlinear interactions among filters. Then, the original outputs from the basic block are rescaled by filter-wise multiplication with the sigmoid activation of the second FC layer of the SE path. We also added residual connections to train a deeper model. The digit in the model name, ReSE-2-Multi, indicates the number of convolution layers in one building block. Finally, we concatenate three hidden layers to take account of different levels of abstraction.



\section{Datasets and Results}

\begin{table*}[t]
\vspace{1.6 mm}
\centering
\vspace{-5mm}
\caption{Description of the three datasets, models and results.}
\label{table1}
\def\arraystretch{1.1}
\resizebox{\textwidth}{!}{\begin{tabular}{llll}
\toprule 
 & \textbf{Music}                                                                                              & \textbf{Speech}                                                                               & \textbf{Scene sound}                                                                                                                                  \\ \midrule
Dataset  &&&\\            & MagnaTagATune (MTAT) \cite{law2009evaluation}                                                                                       & Speech Commands Dataset \cite{speechcommands,WinNT}                                                                                     & DCASE 2017 Task 4 \cite{mesaros2017dcase} \\ & & (TensorFlow Speech Recognition Challenge) &(subtask A)                                                                                                                                     \\ 
Task                 & Music auto-tagging                                                                                               & Speech command recognition                                                                        & Acoustic scene tagging                                                                                                                                \\ 
\# of classes           & 50 tags                                                                                                     & 10 commands + "silence" + "unknown"                                                                                  & 17 sound events   \\ & & (31 classes in training / 12 classes in testing) &                                                                                                                                   \\ 
Labels               & Multi-label                                                                                                 & Multi-class                                                                                  & Multi-label                                                                                                                                           \\ 
Sampling rate& 22,050Hz& 16,000Hz& 44,100Hz \\
Dataset split          & 15,244 / 1529 / 4332                                                                                 & 57,929 / 6798 / 30\% of 158,538 & 45,313 / 5859 / 488                          \\ (train/valid/test)& & (public leaderboard test setting) & (development set) \\                               
Duration          & 29 seconds                                                                                  & 1 second & 10 seconds \\ 
Description          & Collected using the &  Single-word speaking commands, & Subset of AudioSet \cite{gemmeke2017audio}, \\ 
          & TagATune game and music &  rather than conversational sentences  & YouTube clips focusing on  \\ 
          & from Magnatune. &   & vehicle and warning sounds. \\


\midrule
Model  &&&\\  
 & (resampled to 16,000Hz)    &    & (resampled to 16,000Hz) \\ 
Input size & 39,366 samples, 2.46 sec  & 16,000 samples, 1 sec  & 19,683 samples, 1.23 sec  \\ 
 
\# of segments              & 12 segments per clip     & 1 segments per clip   & 9 segments per clip  \\ 
\# of blocks &  9 blocks      & 8 blocks    & 8 blocks   \\

\midrule
Results &&&\\             & AUC      & Accuracy    & F-score (instance-based)   \\

\textbf{SampleCNN}              & 0.9033     & 84\%    & 38.9\%   \\
\textbf{ReSE-2+Multi}              & 0.9091     & 86\%    & 45.1\%   \\

State-of-the-art              & 0.9113 \cite{kim2017rese}    & 88\% (as of Nov 29, 2017) \cite{WinNT}   & 57.7\% \cite{xu2017surrey}  \\

\bottomrule

\end{tabular}}
\end{table*}

We validate the effectiveness of the proposed models on music auto-tagging, speech command recognition and acoustic scene tagging. The details about the datasets for the tasks are summarized in Table \ref{table1}. Note that we resampled all audio samples to 16,000Hz in order to verify how extensible the models are for the three sub-domain of audio tasks in the same condition. However, we configured the input size of the models for each dataset to commonly used size in each domain. Then, we set the number of building blocks according to the input size of the models. We averaged the prediction score for all segments of one clip in testing phase if the input size of the model is shorter than the duration of the clip. 



Table \ref{table1} compares the results from the two sample-level CNN models. The performances are reported using commonly used evaluation metrics for each task. We also compare them to state-of-the-art performance on each dataset. In general, the ReSE-2-Multi model shows close performance to the state-of-the-art results except for the DCASE task. However, in \cite{xu2017surrey}, they used data balancing, ensemble networks and auto thresholding techniques. Without those techniques, they report that their CRNN model achieved 42.0\% F-score value which is lower than our result with ReSE-2-Multi. Also, for the music auto-tagging task on MTAT, the state-of-the-art result was achieved by the ReSE-2-Multi model. In this case, the performance degradation is seen to be caused by downsampling to 16,000Hz. 



\section{Filter Visualizations}

\begin{figure}[t]
\vspace{1.6 mm}
  \centering
  \centerline{\includegraphics[width=\textwidth]{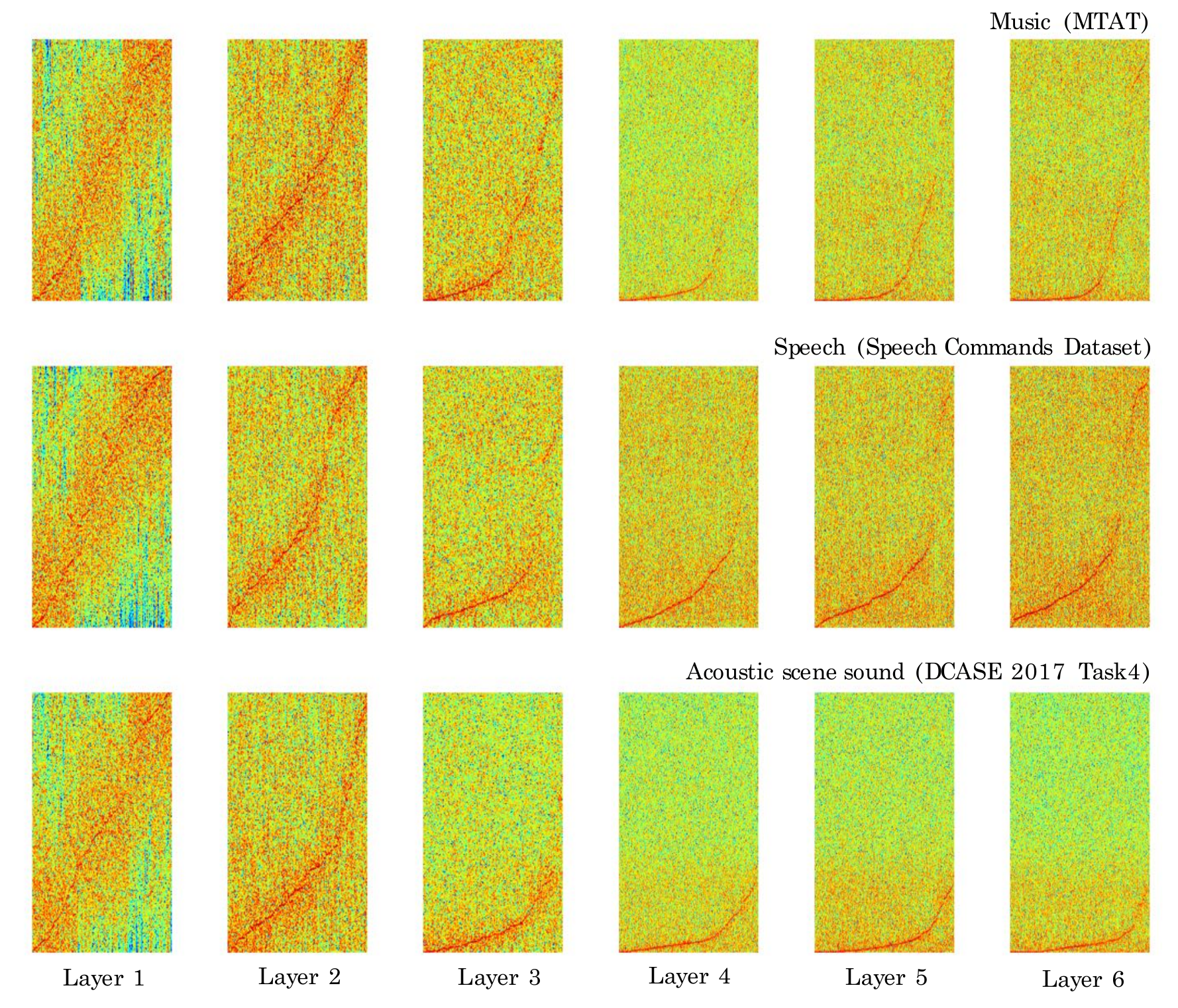}}
  \caption{The spectrum of learned filter estimates for the three datasets in SampleCNN. They are sorted by the frequency at which the magnitude is maximum. The x-axis
represents the index of the filters and the y-axis represents the frequency (ranging from 0 to 8000Hz for all figures). The visualizations were obtained using a gradient ascent method that finds the input waveform that maximizes the activation of a filter at each layer.}
  \label{figure1}
\end{figure}

Visualizing the filters at each layer allows better understanding of representation learning in the hierarchical networks. Since both models yielded similar patterns of learned filters at each layer, we visualize them only for the sampleCNN model.  Figure \ref{figure1} shows the filters obtained by an activation maximization method \cite{erhan2009visualizing}. To show the patterns more clearly, we visualized them as spectrum in the frequency domain and sorted them by the frequency at which the magnitude is maximum \cite{lee2017sample}. In this case, we set the size of the initial random noise to 729 ($=3^6$) samples, so that the estimated filters have typical frame-sized shape which will make the spectrum clearer. Also, for the first 6 layers we only used 3-sized filters and sub-sampling layers. Thus, the temporal dimension of the 6th layer output becomes one in this configuration. For other layers, we averaged remaining temporal dimension so as to make a single activation loss value. Finally, we conducted log-based magnitude compression on the spectrum.

From the figure, we can first find that they are sensitive to log-scaled frequency along the layers, such as mel-frequency spectrogram that is widely used in audio classification tasks. Second, when comparing acoustic scene sound with the other domains, the learned filters tend to have more low-frequency concentration and less complex patterns. This is probably because the DCASE task 4 dataset is made up of simple traffic and warning sounds. Finally, between music and speech, we can observe that more filters explain low-frequency content in music than speech. 


\section{Conclusions}
We presented the two sample-level CNN models that directly take raw waveforms as input and have filters with small granularity. We evaluated them on three audio classification tasks. The results show the possibility that they can be applied to different audio domains as a true end-to-end model. As future work, we will investigate more filter visualization techniques to have better understanding of the models. 


\bibliographystyle{unsrt}
\bibliography{refs}

\end{document}